\documentclass[runningheads]{llncs}

\def\dofullversion{1}

\usepackage[
    lambda, advantage, operators, sets, adversary, landau, probability, notions,
    logic, ff, mm, primitives, events, complexity, asymptotics, keys
]{cryptocode}

\usepackage{mathtools}
\usepackage{amsmath}
\usepackage{amssymb}

\usepackage{graphicx}
\usepackage[position=bottom]{subfig}
\usepackage{tikz}
\usepackage{tikz-qtree}
\usetikzlibrary{shapes.misc,positioning,arrows,snakes}

\usepackage{color}
\usepackage{colortbl}
\definecolor{myteal}{HTML}{009999}
\definecolor{mylime}{HTML}{809933}
\definecolor{mypurple}{HTML}{993380}
\definecolor{myred}{HTML}{E04644}
\definecolor{mydgreen}{HTML}{008000}
\definecolor{mydblue}{HTML}{2809B2}

\usepackage{hyperref}
\ifx\dofullversion\undefined%
	\hypersetup{
	  colorlinks = true,                          
	  urlcolor   = blue,                            
	  linkcolor  = blue,                                  
	  citecolor  = blue,                                       
	}%
\else%
	\hypersetup{
	  colorlinks = true,
	  urlcolor   = mydblue,
	  linkcolor  = mydblue,
	  citecolor  = mydgreen,
	}%
\fi

\usepackage{stackengine}
\newcommand\barbelow[1]{\stackunder[1.2pt]{#1}{\rule{.8ex}{.075ex}}}

\newcommand{\redact}[1]{%
  \ifx\doredact\undefined%
    #1%
  \else%
    \emph{redacted during submission}%
  \fi
}

\newcommand{\fullversion}[2]{%
  \ifx\dofullversion\undefined%
  	#2%
  \else%
    #1%
  \fi
}

\title{%
  Aggregation-Based\\
  Certificate Transparency Gossip
}
\titlerunning{%
  Aggregation-Based Certificate Transparency Gossip
}

\ifx\doredact\undefined%
	\author{%
	  R.\ Dahlberg \and
	  T.\ Pulls \and
	  J.\ Vestin \and
	  T.\ H{\o}iland-J{\o}rgensen \and
	  A.\ Kassler
	}

	\authorrunning{%
	  R.\ Dahlberg \emph{et~al.}
	}

	\institute{%
	  Dept.\ of Mathematics and Computer Science\\
	  Karlstad University, Karlstad, Sweden \\
	  \email{rasmus.dahlberg@kau.se}
	}
\fi

\begin{document}
  \maketitle
  \begin{abstract}
    Certificate Transparency (CT) requires that every CA-issued TLS
certificate must be publicly logged.
While a CT log need not be trusted in theory, it relies on the assumption that
every client observes and cryptographically verifies the same log. As such, some
form of gossip mechanism is needed in practice. Despite CT being adopted by
several major browser vendors, no gossip mechanism is widely deployed.
We suggest an aggregation-based gossip mechanism that passively observes
cryptographic material that CT logs emit in plaintext, aggregating at packet
processors (such as routers and switches) to periodically verify log consistency
off-path. In other words, gossip is provided as-a-service by the network.
Based on 20 days of RIPE Atlas measurements that represent clients from 3500
autonomous systems and 40\% of the IPv4 space, our proposal can be deployed
incrementally for a realistic threat model with significant protection against
split-viewing CT logs. We also show that aggregation-based gossip can be
implemented for a variety of packet processors using P4 and XDP, running at
10~Gbps line-speed.

  \end{abstract}

  \section{Introduction} \label{sec:introduction}
The HTTPS ecosystem is going through a paradigm shift. As opposed to blindly
trusting that Certificate Authorities (CAs) only issue certificates to the
rightful domain owners%
  ---a model known for its weakest-link security~\cite{ca-ecosystem}---%
transparency into the set of issued certificates is incrementally being
required by major browser vendors~\cite{apple-ct,google-ct}.  This transparency
is forced and takes the form of Certificate Transparency (CT) logs:
  the idea is to reject any TLS certificate that have yet to be publicly logged,
  such that domain owners can monitor the logs for client\mbox{-}accepted certificates
	to \emph{detect} certificate mis-issuance \emph{after the fact}~%
	\cite{ct}.
While the requirement of certificate logging is a significant improvement to the
HTTPS ecosystem, the underlying problem of trusting CAs cannot be solved by the
status quo of trusted CT logs (Sect.~\ref{sec:background:ct}). Therefore, it is
paramount that nobody needs to trust these logs once incremental deployments
mature further.

CT is formalized and cryptographically verifiable~\cite{ct-formal}, supporting
inclusion and consistency proofs.
This means that a client can verify whether a log is
operated correctly:
  said certificates are included in the log, and
  nothing is being removed or modified.
Despite the ability to cryptographically verify these two properties, there are
no assurances that everybody observes \emph{the same log}~%
\cite{chuat-gossip,ct}.  For example, certificate mis-issuance would not be
detected by a domain owner that monitors the logs if fraudulently issued
certificates are shown to the clients selectively.  A log that serves
different versions of itself presents a \emph{split view}~\cite{ietf-gossip}.
Unless such log misbehaviour can be detected we must trust it not to happen.

The solution to the split viewing problem is a gossip mechanism which ensures
that everybody observes \emph{the same} consistent log~\cite{ct}.  This
assumption is simple in theory but remarkably hard in practice due to
	client privacy,
	varying threat models, and
	deployment challenges~\cite{ietf-gossip,cosi}.
While Google started on a package that supports
	minimal gossip~\cite{minimal-gossip} and
	the mechanisms of Nordberg \emph{et~al.}~\cite{ietf-gossip},
there is ``next to no deployment in the wild''~\cite{little-or-no-gossip}.
To this end we propose a gossip mechanism that helps detecting split-view
attacks retroactively based on the idea of packet processors such as routers
and middleboxes that \emph{aggregate} Signed Tree Heads (STHs)---succinct
representations of the logs' states---that are exposed to the network \emph{in
plaintext}.
The aggregated STHs are then used to challenge the logs to prove consistency
via an off-path, such that the logs cannot distinguish between challenges that
come from different aggregators.  Given this indistinguishability assumption it
is non-trivial to serve a consistent split-view to an unknown location~%
\cite{mpaudit}.  Thus, all aggregators must be on the same view, and accordingly
all clients that are covered by these aggregators must also be on the same view
\emph{despite not doing any explicit gossip themselves} because gossip is
provided as-a-service by the network.  An isolated client (i.e., untrusted
network path to the aggregator) is notably beyond reach of any retroactive
gossip~\cite{cosi}.

The premise of having STHs in plaintext is controversial given current trends to
encrypt transport protocols, which is otherwise an approach that combats
inspection of network traffic and protocol ossification~\cite{HTTPSintercept,%
TCPoss}.  We argue that keeping gossip related material in plaintext to support
aggregation-based gossip comes with few downsides though:
  it is easy to implement,
  there are no major negative privacy impacts, and
  it would offer significant protection for a large portion of the Internet
    with a realistic threat model \emph{despite relatively small deployment
    efforts}.
The three main limitations are
  no protection against isolated clients,
  reliance on clients that fetch STHs from the logs in plaintext, and
  possible concerns surrounding protocol ossification~\cite{TCPoss}.
Our contributions are as follows.
\begin{itemize}
  \item Design and security considerations for a network-based gossip mechanism
    that passively aggregates STHs to verify log consistency off-path
    (Sect.~\ref{sec:design}).
  \item Generic implementations of the aggregation step using P4~\cite{p4} and
    XDP~\cite{xdp} for plaintext STHs, supporting line-speed packet
    processing on systems that range from switches, routers, network interface
    cards, and Linux (Sect.~\ref{sec:implementation}).
  \item A simulation based on RIPE Atlas measurements that evaluate the impact
    of deploying aggregation-based gossip at ASes and IXPs. Our evaluation shows
    that incremental roll-out at well-connected locations would protect a
    significant portion of all Internet clients from undetected split views
    (Sect.~\ref{sec:measurements}).
\end{itemize}

Besides the sections referenced above the paper first introduces necessary
background (Sect.~\ref{sec:background}) and finally provides discussion and
conclusions (Sect.~\ref{sec:related}--\ref{sec:conclusion}).
Appendices~\ref{app:implementation}--\ref{app:data} provide further
implementation and public data set details.

  \section{Background} \label{sec:background}
\fullversion{%
	First additional prerequisites are provided on CT and the status quo,
	then the techniques which allow us to program custom packet processors are
	introduced.%
}{}

\subsection{Certificate Transparency} \label{sec:background:ct}
The main motivation of CT is that the CA ecosystem is error-prone~\cite{laurie}:
	a CA can normally issue certificates for \emph{any} domain name, and
	given that there are hundreds of trusted CAs an attacker only needs to
	target the weakest link~\cite{ca-ecosystem}. 
While the requirement of CT logging all certificates cannot prevent mis-issuance
proactively, it allows anyone to detect it retroactively by monitoring the logs%
~\cite{ct}.  After a log promises to include a certificate by issuing a Signed
Certificate Timestamp (SCT), a new STH including the appended certificate
must be issued within a Maximum Merge Delay (MMD). Typically, logs use 24~hour
MMDs. Should non-included SCTs and/or inconsistent STHs be found,
binding evidence of log misbehaviour exists because these statements are
digitally signed.  Other than MMD a log's policy defines parameters such as STH
frequency:
	the number of STHs that can be issued during an MMD, making it harder to
	track clients~\cite{ietf-gossip}.

CT is being deployed across Apple's platform~\cite{apple-ct} and Google's
Chrome~\cite{google-ct}. The status quo is to trust a CA-signed certificate if
it is accompanied by two or more SCTs, thereby relying on at least one log to
append each certificate so that mis-issuance can be detected by monitors that
inspect the logs. The next step of this incremental deployment is to
\emph{verify} that these certificates are actually logged by querying for
inclusion~\cite{google-gossip}, and that the log's append-only property is
respected by challenging the log to prove STH consistency. Finally, to fully
distrust CT logs we need mechanisms that detect split-views.  One such mechanism
which is based on programmable packet processors (introduced next) is presented
in Sect.~\ref{sec:design}, and it is compared to related work on CT gossip in
Sect.~\ref{sec:related}.

\subsection{Programmable Data Planes} \label{sec:background:pdp}
Packet processors such as switches, routers, and network interface cards
are typically integrated tightly using customized hardware and application-%
specific integrated circuits. This inflexible design limits the
potential for innovation and leads to long product upgrade cycles, where it
takes \emph{years} to introduce new processing capabilities and support for
different protocols and header fields (mostly following lengthy
standardization cycles).
The recent shift towards flexible \emph{match+action} packet-processing
pipelines---including
  RMT~\cite{rmt},
  Intel Flexpipe\footnote{%
		\fullversion
			{\url{https://www.intel.com/content/dam/www/public/us/en/documents/product-briefs/ethernet-switch-fm6000-series-brief.pdf} (n.d.)}
			{\url{https://www.intel.com/content/dam/www/public/us/en/documents/product-briefs/ethernet-switch-fm6000-series-brief.pdf}}
  },
  Cavium XPA\footnote{%
	  \fullversion
		{\url{https://web.archive.org/web/20170707175037/https://cavium.com/newsevents-cavium-and-xpliant-introduce-a-fully-programmable-switch-silicon-family.html} (n.d.)}
		{\url{https://cavium.com/newsevents-cavium-and-xpliant-introduce-a-fully-programmable-switch-silicon-family.html}}
  }, and
  Barefoot Tofino\footnote{%
	  \fullversion
		{\url{https://web.archive.org/web/20180105002028/https://barefootnetworks.com/products/brief-tofino/} (n.d.)}
		{\url{https://barefootnetworks.com/products/brief-tofino/}}
  }---%
now have the potential to change the way in which packet processing hardware is
implemented:
  it enables programmability using high-level languages such as P4 (see below),
  while at the same time maintaining performance comparable to fixed-function
  chips.

\subsubsection{P4.}
The main goal of P4 is to simplify
  \barbelow{p}rogramming of
  \barbelow{p}rotocol-independent
  \barbelow{p}acket
  \barbelow{p}rocessors
by providing an abstract programming model for the network data plane~\cite{p4}.
In this setting the functionality of a packet processing device is specified
without assuming any hardwired protocols and headers. Consequently, a P4 program
must parse headers and connect the values of those protocol fields to the
actions that should be executed based on a pipeline of reconfigurable
match+action tables.
Based on the specified P4 code, a front-end compiler generates a high-level
intermediate representation that a back-end compiler uses to create a target-%
dependent program representation. Compilers are available for several platforms,
including
  the software-based simple switch architecture\footnote{%
	  \fullversion
		{\url{https://github.com/p4lang/p4c-bm} (2018)}
		{\url{https://github.com/p4lang/p4c-bm}}
  },
  SDNet for Xilinx NetFPGA boards~\cite{p4netfpga}, and
  Netronome's smart network interfaces~\cite{p4netronome}.
It is also possible to compile basic P4 programs into eBPF byte code.%
\footnote{%
	\fullversion
		{\url{https://github.com/iovisor/bcc/tree/master/src/cc/frontends/p4} (2018)}
		{\url{https://github.com/iovisor/bcc/tree/master/src/cc/frontends/p4}}
}

\subsubsection{XDP.}
The Berkeley Packet Filter (BPF) is a Linux-based packet filtering mechanism~%
\cite{bpf}. Verified bytecode is injected from user space, and executed for each
received packet in kernel space by a just-in-time compiler. Extended BPF (eBPF)
enhances the original BPF concept, enabling faster runtime and many new
features.\footnote{%
	\fullversion
		{\url{https://github.com/netoptimizer/prototype-kernel/blob/master/kernel/Documentation/bpf/index.rst} (2017)}
		{\url{https://github.com/netoptimizer/prototype-kernel/blob/master/kernel/Documentation/bpf/index.rst}}
} For example, an eBPF program can be attached to the Linux traffic control
tool \texttt{tc}, and additional hooks were defined for a faster eXpress
Data Path (XDP)~\cite{xdp}. In contrast to the Intel Data Plane Development Kit
(DPDK) which runs in user space and completely controls a given network
interface that supports a DPDK driver,\footnote{%
	\fullversion
		{\url{https://web.archive.org/web/20180520162550/https://dpdk.org/} (n.d.)}
		{\url{https://dpdk.org/}}
} XDP cooperates with the Linux stack to achieve fast, programmable, and
reconfigurable packet processing using C-like programs.

  \section{Design} \label{sec:design}
An overview of aggregation-based gossip is shown in Fig.~\ref{fig:agg}. The
setting consists of logs that send plaintext STHs to clients over a network, and
as part of the network inline \emph{packet processors} passively aggregate
observed STHs to their own off-path \emph{challengers} which challenge the logs
to prove consistency. A log cannot present split views to different clients that
share an aggregating vantage point because it would trivially be detected by
that vantage point's challenger. A log also cannot present a persistent split
view to different challengers because they are off-path in the sense that they
are indistinguishable from one another. This means that every client that is
covered by an aggregator must be on the same view because at least one
challenger will otherwise detect an inconsistency and report it. A client that
is not directly covered by an aggregator may receive indirect protection in the
form of herd immunity. This is discussed in Sect.~%
\ref{sec:discussion:herd}.
\ifx\dofullversion\undefined\begin{figure}[h]\else\begin{figure}[!t]\fi
  \centering
  \includegraphics[width=.8\columnwidth]{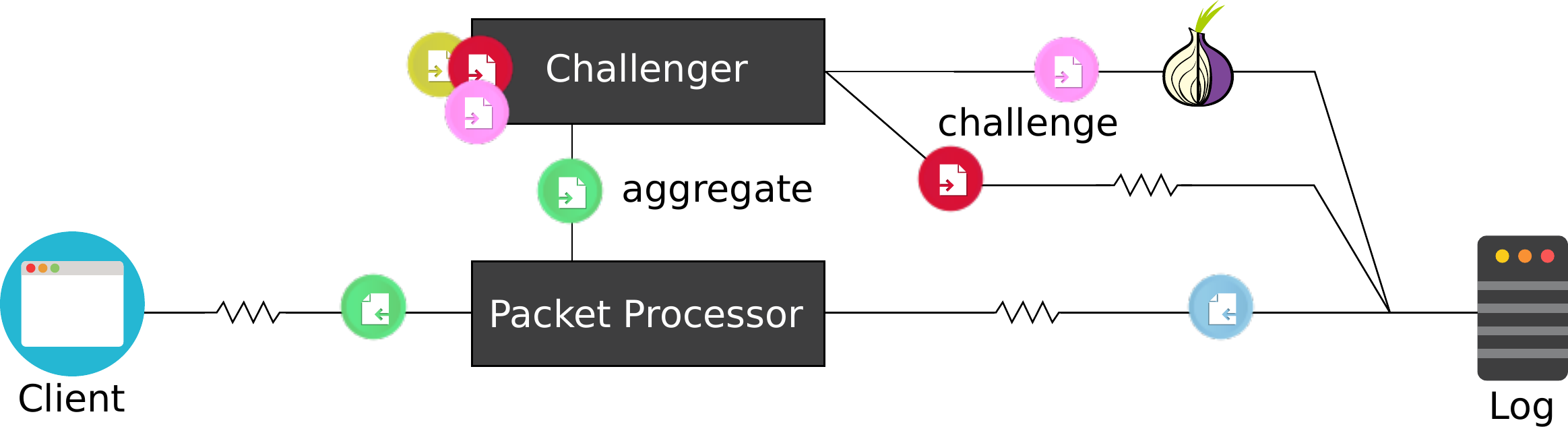}
  \caption[ ]{%
    Packet processor that aggregates plaintext STHs for off-path verification.
  }
  \label{fig:agg}
\end{figure}

\subsection{Threat Model and Security Notion} \label{sec:agg:thr}
The overarching threat is undetectable domain impersonation (ex-post) by an
attacker that is capable of compromising at least one CA and a sufficient number
of CT logs to convince a client into accepting a forged certificate.
We assume that any illegitimately issued certificate
would be detected by the legitimate domain owner through self or delegated
third-party monitoring.
This means that an attacker must either provide a split view towards the victim
or the monitoring entity.
We also assume that clients query the logs for certificate inclusion based on
STHs that it acquires from the logs via plaintext mechanisms that packet
processors can observe, and that some other entities than challengers process
STHs using the chosen off-paths (Sect~\ref{sec:discussion:limitations}).
We do not account for the fact that CA compromises may be detected by other
means, focusing solely on split views.

\subsubsection{Limitations.}
Our gossip mechanism is limited to STHs that packet processors can observe.
As such, a client isolated by an attacker is not protected. We limit ourselves
to attackers that act over a network some distance
(in the sense of network path length) from a client in plaintext so that
aggregation can take place. Our limitations and assumptions are further
discussed in Sect.~\ref{sec:discussion:limitations}.

\subsubsection{Attackers.}
Exceptionally powerful attackers can isolate clients, \emph{but clients are not
necessarily easy to isolate} for a significant number of relevant attackers.
Isolation may require physical control over a device,\footnote{%
  \fullversion
	{See FBI-Apple San Bernardino case: \url{https://web.archive.org/web/20180520135200/https://www.eff.org/cases/apple-challenges-fbi-all-writs-act-order} (2016)}
	{\url{https://www.eff.org/cases/apple-challenges-fbi-all-writs-act-order}}
}
clients may be using anonymity networks like Tor where path selection is
inherently unpredictable~\cite{tor}, or sufficiently large parts of the network
cannot be controlled to ensure that no aggregation takes place.
This may be the case if we consider
  a nation state actor attacking another nation state actor,
  the prevalence of edge security middleboxes, and
  that home routers or NICs nearby the clients could aggregate.
Any attacker that cannot account for these considerations is within our
threat model.

\subsubsection{Security Notion.}
To bypass our approach towards gossip an adaptive attacker may attempt to
actively probe the network for aggregating packet processors. This leads us to
the key security notion:
  \emph{aggregation indistinguishability}.
An attacker should not be able to determine if a packet processor is aggregating
STHs. The importance of aggregation indistinguishability motivates the design of
our gossip mechanism into two distinct components:
  aggregation that takes place inline at packet processors, and
  periodic off\mbox{-}path log challenging that checks whether the observed STHs
    are consistent (if not report the misbehaving log).

\subsection{Packet Processor Aggregation} \label{sec:aggregator}
An aggregating packet-processor determines for each packet if it is STH-related.
If so, the packet is cloned and sent to a challenging component for off-path
verification (see Sect.~\ref{sec:challenger}).
The exact definition of \emph{STH-related} depends on the plaintext source, but it
is ultimately the process of inspecting multiple packet headers such as
transport protocol and port number. It should be noted that the original packet
must not be dropped or modified. For example, an aggregator would have a
trivial aggregation distinguisher if it dropped any malformed STH.

For each aggregating packet processor we have to take IP fragmentation into
consideration. Recall that IP fragmentation usually occurs when a packet is
larger than the MTU, splitting it into multiple smaller IP packets that are
reassembled at the destination host. Normally an STH should not be fragmented
because it is much smaller than the de-facto minimum MTU of (at least) 576~%
bytes~\cite{min-mtu,ipv6}, but an attacker could use fragmentation to
\emph{intentionally} spread expected headers across multiple packets.
Assuming stateless packet processing, an aggregator cannot identify
such fragmented packets as STH-related because some header would be absent
  (cf.\ stateless firewalls).
All tiny fragments should therefore be aggregated to account for intentional IP
fragmentation, which appears to have little or no impact on normal traffic
because tiny fragments are anomalies~\cite{frag-study-02}. The threat of 
multi-path fragmentation is discussed in Sect.~\ref{sec:discussion:limitations}.

Large traffic loads must also be taken into account. If an aggregating
packet processor degrades in performance as the portion of STH-related traffic
increases, a distant attacker may probe for such behaviour to determine if a
path contains an aggregator. Each \emph{implementation} must therefore be
evaluated individually for such behaviour, and if trivial aggregation
distinguishers exist this needs to be solved. For example, STH-related traffic
could be aggregated probabilistically to reduce the amount of work
	(Appendix~\ref{app:imp:poc}).
Another option is to load-balance the traffic before aggregation, i.e., avoid
worst-case loads that cannot be handled.

\subsection{Off-Path Log Challenging} \label{sec:challenger}
A challenger is setup to listen for aggregated traffic, reassembling IP
fragments and storing the aggregated STHs for periodic off-path verification.
Periodic off\mbox{-}path verification means that the challenger challenges the log
based on its own (off-path fetched) STHs and the observed (aggregated) STHs to
verify log consistency periodically, e.g., every day.
The definition of \emph{off-path} is that the challenger must not be linkable to
its aggregating packet processor(s) or any other challenger (including itself).
Without an off-path there is no gossip step amongst aggregator-challenger
instances that are operated by different actors, and our approach towards gossip
would only assert that clients behind the same vantage point observe the same
logs. If a log cannot distinguish between different challengers due to the
use of off-paths, however, it is non-trivial to maintain a targeted split-view
towards an unknown location. Therefore, we get a form of \emph{implicit gossip}%
~\cite{mpaudit} because at least one challenger would detect an inconsistency
unless everybody observes the same log. If every challenger observes the same
log, so does every client that is covered by an aggregating packet processor.
Notably the challenger component \emph{does not run inline} to avoid timing
distinguishers. Sect.~\ref{sec:discussion:limitations} discusses aggregation
distinguishers based on unique STH probes.

  \section{Implementation and Distinguishability Experiments} \label{sec:implementation}
There are many different ways to implement the aggregation step. We decided to
use P4 and XDP because a large variety of programmable
packet processors support these languages (Sect~\ref{sec:background:pdp}).
The aggregated plaintext source is assumed to be CT-over-DNS~\cite{ct-over-dns},
which means that a client obtains STHs by fetching IN TXT resource records.
Since languages for programmable packet processors are somewhat restricted,
we facilitated packet processing by requiring that at most one STH is sent per
UDP packet (Appendix~\ref{app:imp:pt}).
This is reasonable because logs should only have one \emph{most recent} STH.
A DNS STH is roughly 170~bytes without any packet headers and
should normally not be fragmented, but to ensure that we do not miss any
intentionally fragmented STHs we aggregate every tiny fragment. We did not
implement the challenging component because it is relatively easy given
an existing off-path. Should any scalability issue arise for the challenger
there is nothing that prevents a distributed front-end that processes the
aggregated material before storage. Storage is not an issue because there are
only a limited amount of unique STHs per day and log
  (one new STH per hour is a common policy, and browsers recognize $\approx 40$
  logs).
Further details can be found on GitHub
  (\redact{\url{https://github.com/rgdd/ctga}})
and in Appendix~\ref{app:implementation}.

\subsubsection{Setup.}
We used a test-bed consisting of
  a traffic generator,
  a traffic receiver, and
  an aggregating target in between.
The first target is a P4-enabled NetFPGA SUME board that runs an adapted version
of our P4 reference implementation.
The second target is a net-next kernel v4.17.0-rc6 Linux machine that runs XDP
on one core with
  a 10~Gb SFP+ X520 82599ES Intel card,
  a $3.6$~GHz Intel Core i7-4790 CPU, and
  16~GB of RAM at 1600~MHz (Hynix/Hyundai).
We would like to determine whether there are any aggregation distinguishers as
the fraction of STHs (experiment 1) and tiny fragments (experiment 2) in the
traffic is increased from 0--100\%, i.e., does performance degrade as a
function of STH-related rate? Non-fragmented STH packets are
411~bytes,\footnote{%
  We used excessively large DNS headers to maximize the packet parsing overhead.
} and tiny fragments are 64~bytes. All background traffic have the same
packet sizes but is not deemed STH-related.

\subsubsection{Results.}
Fig.~\ref{fig:perf-p4} shows throughput as a function of STH-related rate for
the P4-enabled NetFPGA. While we were unable to observe any distinguisher between
normal routing and the edge case of 100\% aggregation for
non-fragmented STH packets, there is a small constant throughput difference for
tiny fragments ($7.5$~Kbps). This is a non-negligible \emph{program
distinguisher} if a packet processor is physically isolated as in our benchmark,
i.e., something other than a routing program is running but it is not
necessarily an aggregator because performance does not degrade as a function
of increased STH-related rate. However, we found such degradation behaviour for the
single-core XDP case (Fig.~\ref{fig:perf-xdp}). If line-speed is higher than
2~Gbps, traffic could be load-balanced to overcome this issue.
\begin{figure}[h]
  \centering
  \ifx\dofullversion\defined\vspace{-.7cm}\fi
  \subfloat[][P4 NetFPGA]{%
    \includegraphics[width=0.5\columnwidth]{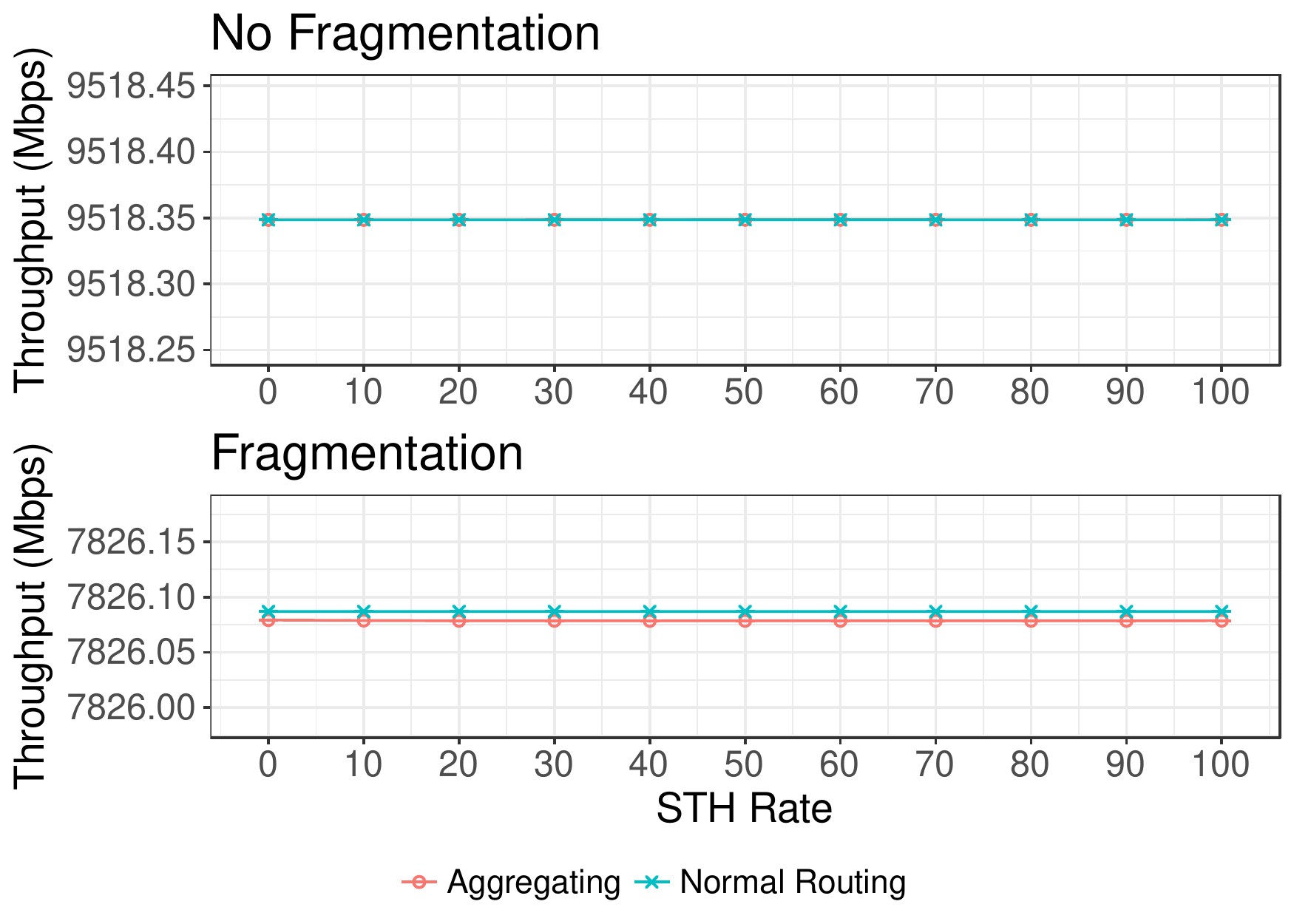}
    \label{fig:perf-p4}
  }
  \subfloat[][XDP on a single core]{%
    \includegraphics[width=0.5\columnwidth]{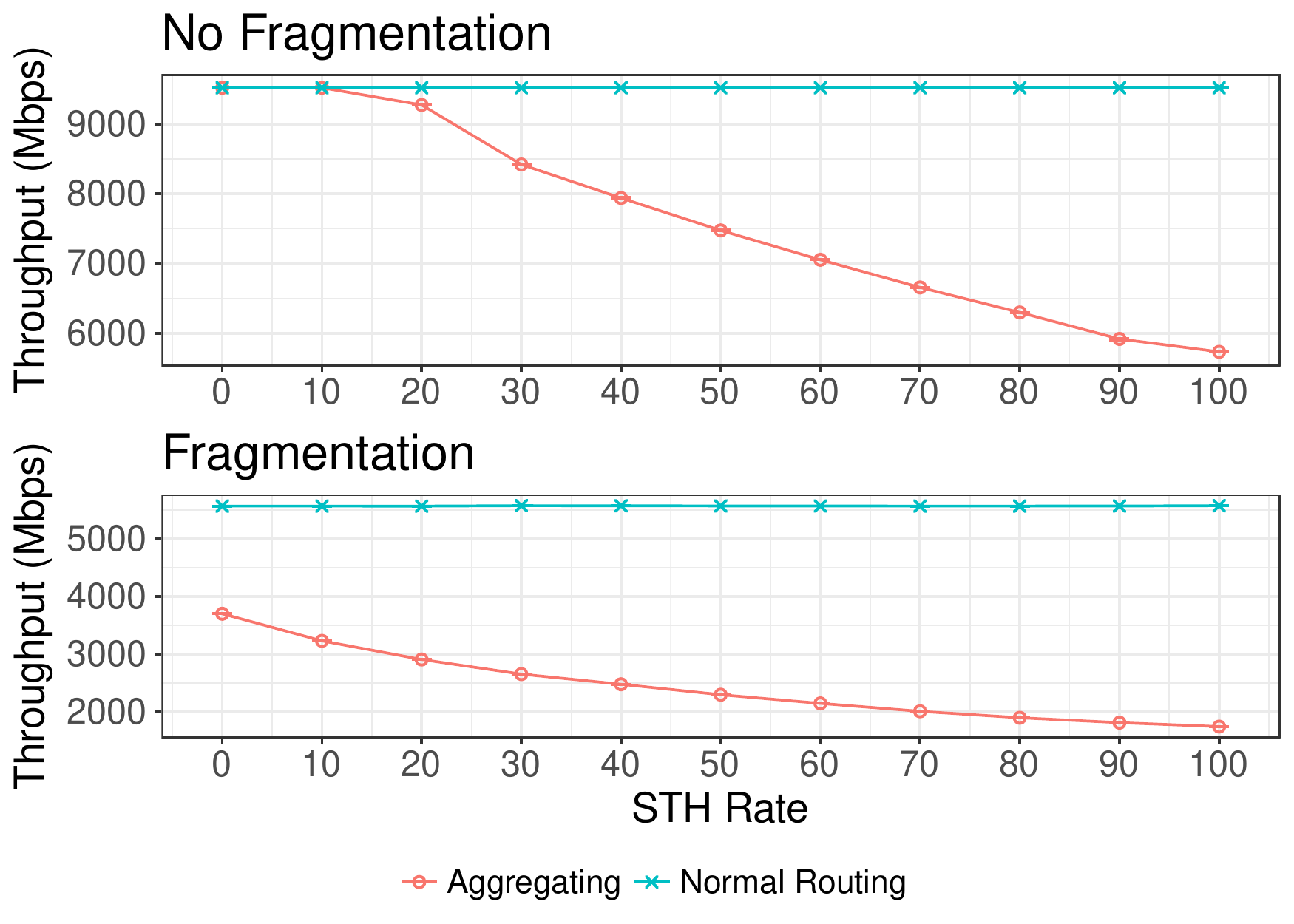}
    \label{fig:perf-xdp}
  }
  \caption{%
    Throughput as a function of STH-related traffic that is aggregated.
  }
  \label{fig:perf}
  \ifx\dofullversion\defined\vspace{-1cm}\fi
\end{figure}

\subsubsection{Lessons learned.}
Aggregation indistinguishability is provided by P4-NetFPGA. For XDP it depends
on the scenario: what is the line-rate criteria and how many cores are
available. Five cores support 10~Gbps aggregation indistinguishability.

  \section{Estimated Impact of Deployment} \label{sec:measurements}
We conducted 20 daily traceroute measurements during spring 2018 on the RIPE
Atlas platform to evaluate the effectiveness of aggregation-based gossip. The
basic idea is to look at client coverage as central ASes and IXPs aggregate
STHs. If any significant client coverage can be achieved, the likelihood of
pulling off an undetected split-view will be small given aggregation
indistinguishability.

\subsubsection{Setup.}
We scheduled RIPE Atlas measurements from roughly 3500 unique ASes that
represent 40\% of the IPv4 space, trace-routing Google's authoritative
CT-over-DNS server and NORDUnet's CT log to simulate clients that fetch DNS STHs
in plaintext (Appendix~\ref{app:data:our}). Each traceroute result is a list of
traversed IPs, and it can be translated into the corresponding ASes and IXPs
using public data sets (Appendix~\ref{app:data:pub}).
In other words, traversed ASes and IXPs can be determined for each probe. Since
we are interested in client coverage as ASes and IXPs aggregate, each
probe is weighted by the IPv4 space of its AS. While an IP address is an
imperfect representation of a client, e.g., an IP may be unused or reused, it
gives a decent idea of how significant it is to cover a given probe.

\subsubsection{Results.}
Fig.~\ref{fig:pl} shows AS/IXP path length and stability from the probes to
the targets.
If the AS path length is one, a single AS is traversed \emph{before reaching the
target}. It is evident that an AS path tends to be one hop longer
towards NORDUnet than Google because there is a rough off-by-one offset on the
x-axis.
A similar trend of greater path length towards NORDUnet can be observed for
IXPs. For example,
  74.0\% of all paths traversed no IXP towards Google, but
  58.5\% of all paths traversed a single IXP towards NORDUnet.
These results can be explained by infrastructural differences of our targets:
  since Google is a worldwide actor an average path should be shorter than
  compared to a region-restricted actor like NORDUnet.
We also observed that AS and IXP paths tend to be quite stable over 20~days
	(the duration of our measurements).
In other words, if AS $a$ and $b$ are traversed it is unlikely to suddenly be
routed via AS~$c$.
\begin{figure}[h]
  \centering
  \includegraphics[width=0.5\columnwidth]{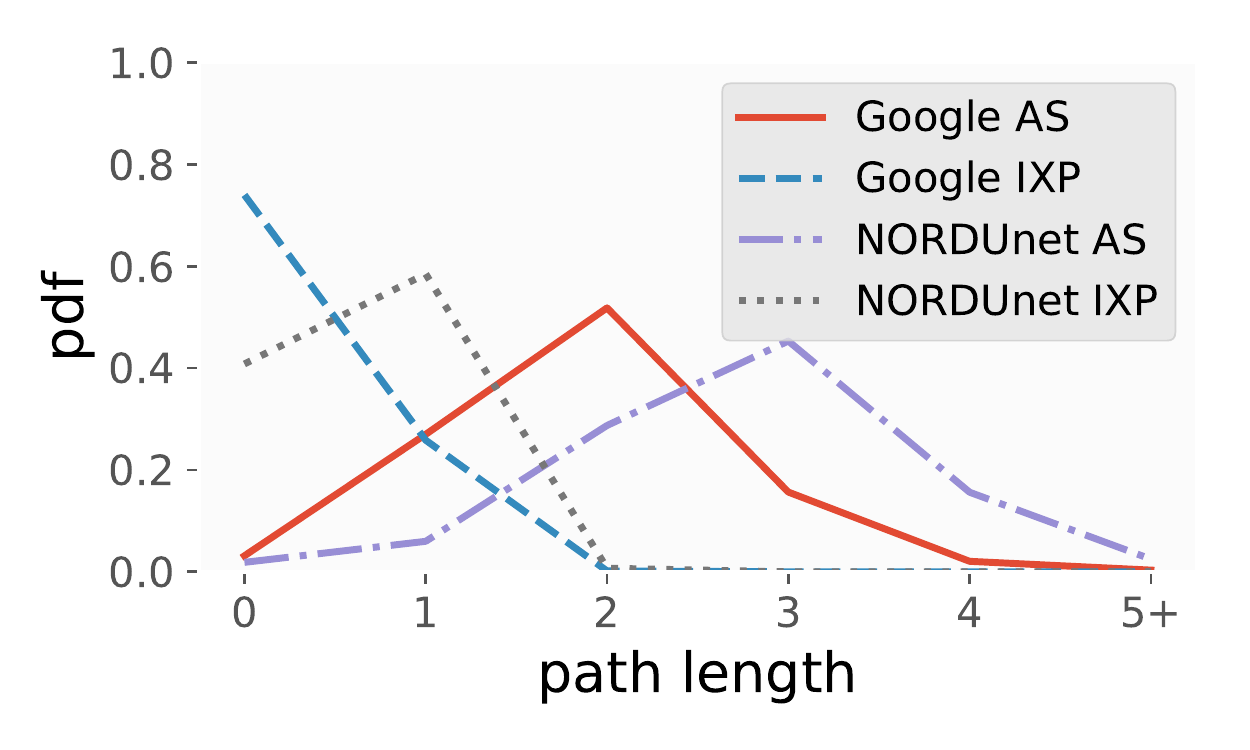}%
  \includegraphics[width=0.5\columnwidth]{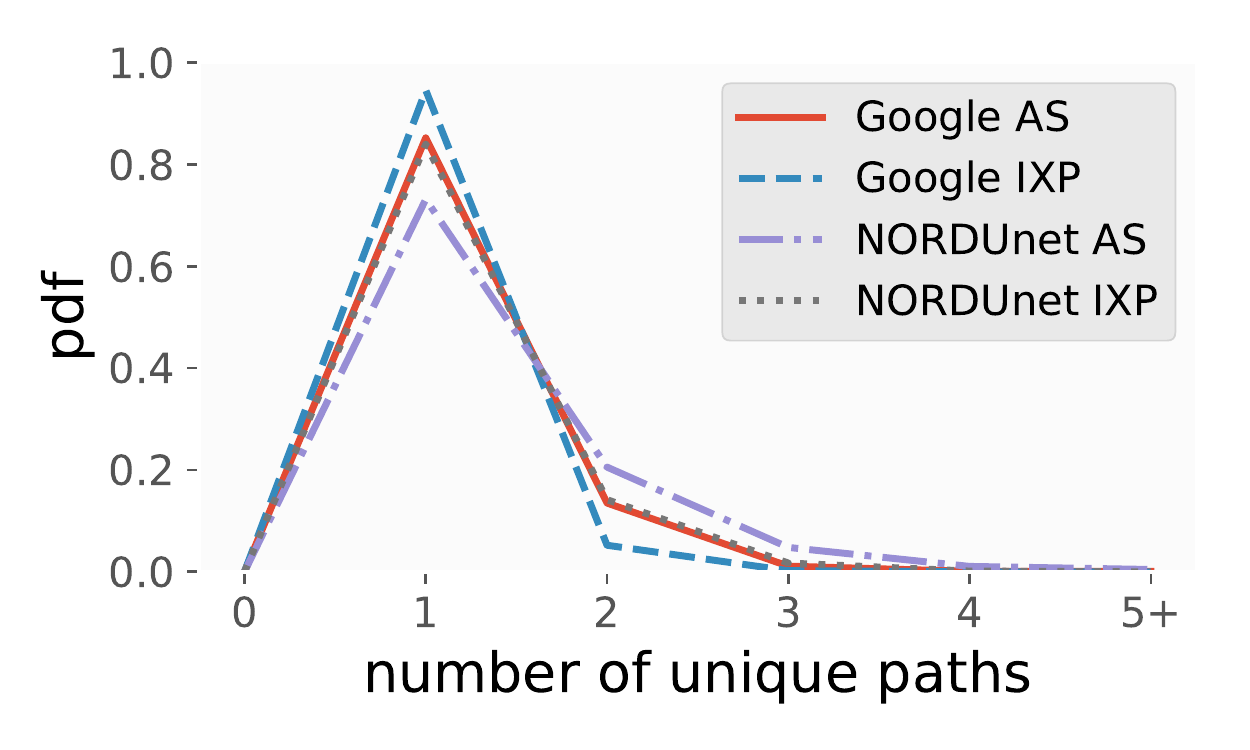}
  \caption{%
    Path length and stability towards Google and NORDUnet.
  }
  \label{fig:pl}
  \ifx\dofullversion\defined\vspace{-.5cm}\fi
\end{figure}

Fig.~\ref{fig:wcov} shows coverage of the RIPE Atlas network as $1...n$ actors
aggregate STHs. For example, 100\% and 50\% coverage means that at least 40\%
and 20\% of the full IPv4 space is covered. The aggregating ASes and IXPs were
selected based on the most commonly traversed vantage points in
our measurements (Pop), as well as CAIDA's largest AS ranking.\footnote{%
	CAIDA ranks ASes based on collected topological data sets (Appendix~%
	\ref{app:data:pub}).
}
We found that more coverage is achieved when targeting
NORDUnet than Google. This is expected given that the paths tend to be longer.
Further, if CAIDA's top-32 enabled aggregation the coverage would be significant
towards Google (31.6\%) and NORDUnet (58.1\%).

\ifx\dofullversion\undefined\begin{figure}[h]\else\begin{figure}[!t]\fi
  \ifx\dofullversion\defined\vspace{-.3cm}\fi
  \centering
  \includegraphics[width=0.5\columnwidth]{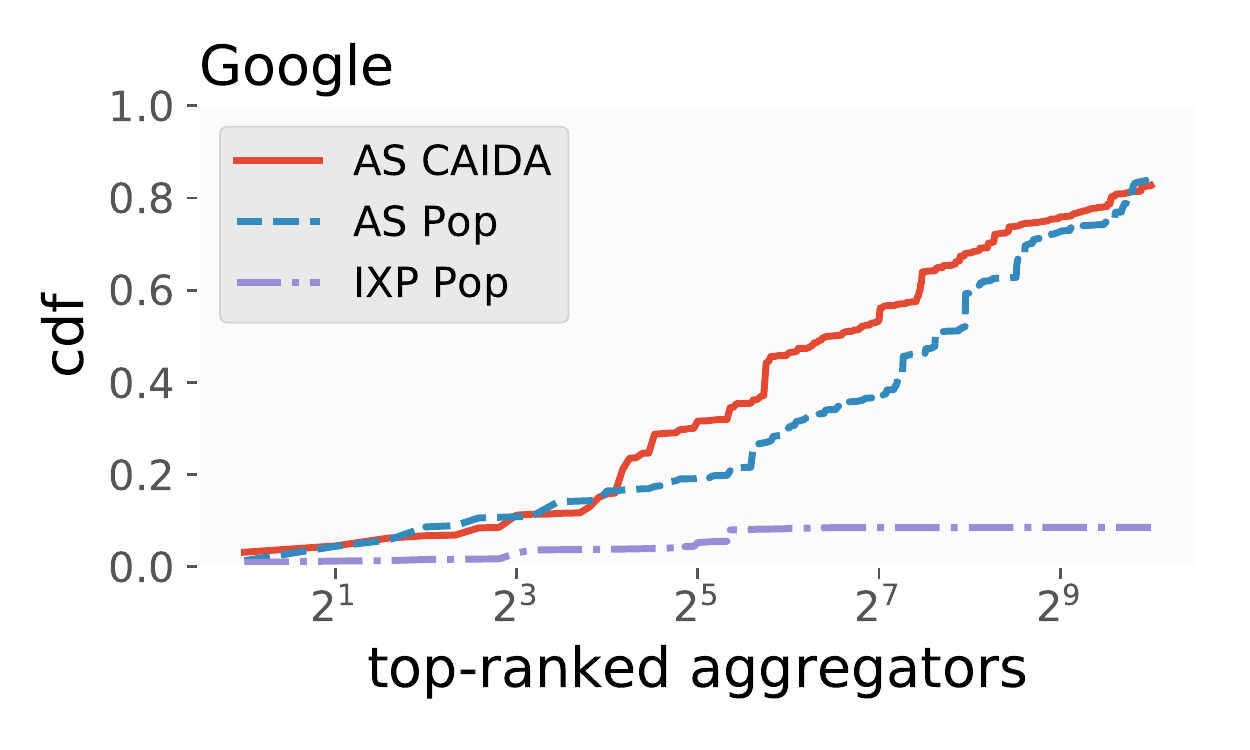}%
  \includegraphics[width=0.5\columnwidth]{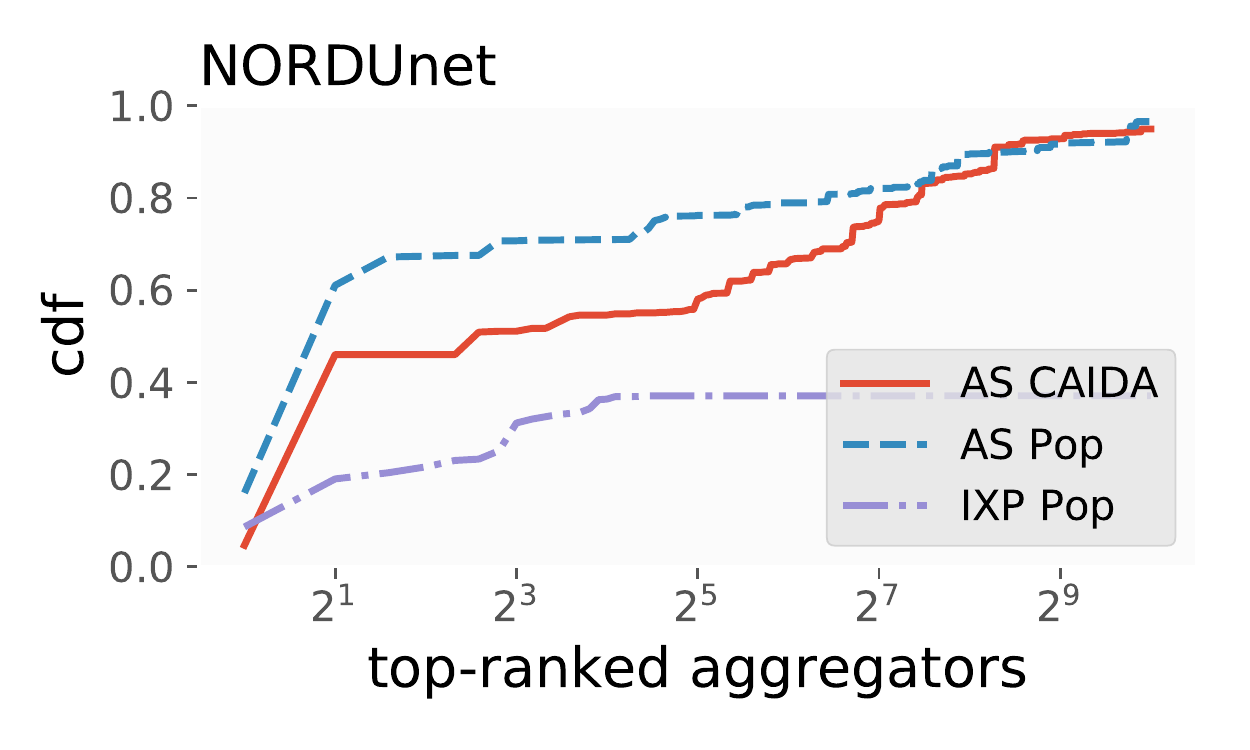}
  \ifx\dofullversion\defined\vspace{-.5cm}\fi
  \caption{%
    Coverage as a function of aggregation opt-in.
  }
  \label{fig:wcov}
  \ifx\dofullversion\defined\vspace{-.5cm}\fi
\end{figure}

\subsubsection{Lessons learned.}
A vast majority of all clients traverse \emph{at least} one AS that could
aggregate. It is relatively rare to traverse IXPs towards Google but not
NORDUnet. We also learned that paths tends to be stable, which means that the
time until split view detection would be at least 20 days \emph{if} it is
possible to find an unprotected client. This increases the importance of
aggregation indistinguishability.
Finally, we identified vantage points that are commonly traversed using Pop, and
these vantage points are represented well by CAIDA's independent AS ranking.
Little opt-in from ASes and IXPs provides significant coverage against an
attacker that is relatively close to a client
  (cf.\ world-wide infrastructure of Google).
Although we got better coverage for NORDUnet, any weak attacker would approach
Google's coverage by renting infrastructure nearby.
Any weak attacker could also circumvent IXP aggregation by detecting the IXP
itself. As such, aggregating at top-ranked ASes should give the best split-view
protection.

  \section{Related Work} \label{sec:related}
Earlier approaches towards CT gossip are categorized as \emph{proactive} or
\emph{retroactive} in Fig.~\ref{fig:related}. We consider an approach proactive
if gossip takes place \emph{before} SCTs and/or STHs reach the broader audience
of clients.
Syta \emph{et~al.} proposed proactive witness cosigning, in which an STH is
collectively signed by a \emph{large} number of witnesses and at most a fraction
of those can be faulty to ensure that a benevolent witness observed an STH~%
\cite{cosi}.
STH cross-logging~\cite{minimal-gossip,ietf-cross-logging,hof-cross-logging} is
similar in that an STH must be proactively disclosed in
another transparency log to be trusted, avoiding any additional cosigning
infrastructure at the cost of reducing the size and diversity of the witnessing
group.
Tomescu and Devadas~\cite{catena} suggested a similar cross-logging scheme,
but split-view detection is instead reduced to the difficulty of forking the
Bitcoin blockchain
  (big-O cost of downloading all block headers as a TLS client).
The final proactive approach is STH pushing, where a trusted third-party
pushes the same verified STH history to a base of clients~\cite{google-gossip}.
\ifx\dofullversion\undefined\begin{figure}[h]\else\begin{figure}[!t]\fi
  \ifx\dofullversion\defined\vspace{-.75cm}\fi
  \centering
  \resizebox{\columnwidth}{!}{%
  \begin{tikzpicture}[%
    ns/.style = {
      draw=none,
    },
    ps/.style = {
      draw,
      -latex,
    },
  ]
    \node[ns](gossip){};
    \node[ns,right=0pt of gossip](retroactive){\textbf{Retroactive}};
    \node[ns,left=0pt of gossip](proactive){\textbf{Proactive}};

    \node[ns,left=12pt of proactive](cross){STH cross-logging~\cite{minimal-gossip,ietf-cross-logging,hof-cross-logging,catena}};
    \node[ns,above=0pt of cross](push){STH pushing~\cite{google-gossip}};
    \node[ns,below=0pt of cross](cosi){STH cosigning~\cite{cosi}};

    \path[ps] (proactive) -- (push.east);
    \path[ps] (proactive) -- (cross);
    \path[ps] (proactive) -- (cosi.east);

    \node[ns,right=12pt of retroactive](implicit){Implicit via multipath~\cite{mpaudit}};
    \node[ns,above=0pt of implicit](pool){STH pooling~\cite{chuat-gossip,ietf-gossip}};
    \node[ns,below=0pt of implicit](trust){Trusted auditing~\cite{ietf-gossip}};
    \node[ns,above=14pt of retroactive.north east](feedback){SCT feedback~\cite{ietf-gossip}};
    \node[ns,below=14pt of retroactive.south east](bee){CT honey bee~\cite{ct-honey-bee}};

    \path[ps] (retroactive) -- (feedback);
    \path[ps] (retroactive) -- (pool.west);
    \path[ps] (retroactive) -- (implicit);
    \path[ps] (retroactive) -- (trust.west);
    \path[ps] (retroactive) -- (bee);
  \end{tikzpicture}
}
  \ifx\dofullversion\defined\vspace{-.6cm}\fi
  \caption{%
    A categorization of approaches towards CT gossip.
  }
  \label{fig:related}
  \ifx\dofullversion\defined\vspace{-.6cm}\fi
\end{figure}
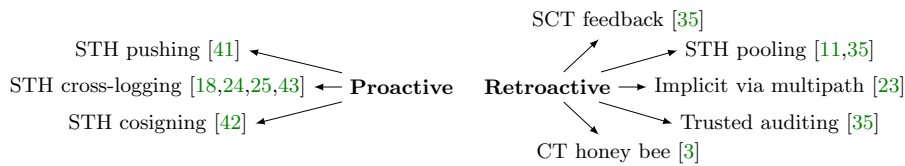

We consider a gossip mechanism retroactive if gossip takes place \emph{after}
SCTs and/or STHs reach the broader audience of clients.
Chuat \emph{et~al.} proposed that TLS clients and TLS servers be modified to
pool exchanged STHs and relevant consistency proofs~\cite{chuat-gossip}.
Nordberg \emph{et~al.} continued this line of work, suggesting privacy-%
preserving client-server pollination of fresh STHs~\cite{ietf-gossip}. Nordberg
\emph{et~al.} also proposed that clients feedback SCTs and certificate chains on
every server revisit, and that trusted auditor relationships could be engaged if
privacy need not be protected.
The latter is somewhat similar to the formalized client-monitor gossip of
Chase and Meiklejohn~\cite{transparency-overlays}, as well as the CT honey bee
project where a client process fetches and submits STHs to a pre-%
compiled list of auditors~\cite{ct-honey-bee}.
Laurie suggested that a client can resolve privacy-sensitive SCTs to privacy-%
insensitive STHs via DNS (which are easier to gossip)~\cite{ct-over-dns}.
Private information retrievals could likely achieve something similar~%
\cite{ct-pir}.
Assuming that TLS clients are indistinguishable from one another, split-view
detection could also be implicit as proposed by Gunn \emph{et~al.} for the
verifiable key-value store CONIKS~\cite{mpaudit,coniks}.

Given that aggregation-based gossip takes place after an STH is issued, it is a
retroactive approach. As such, we cannot protect an isolated client from split-%
views~\cite{cosi}. Similar to STH pooling and STH pollination, we rely on
client-driven communication and an existing infrastructure of packet processors
to aggregate.
Our off-path verification is
based on the same multi-path probing and indistinguishability assumptions as
Gunn \emph{et~al.}~\cite{doublecheck,mpaudit,perspectives}. Further, given that
aggregation is application neutral and deployable on hosts, it could provide
gossip \emph{for} the CT honey bee project (assuming plaintext STHs) and any
other transparency application like Trillian~\cite{vds}. Another benefit when
compared to browsing-centric and vendor-specific approaches is that a plethora
of HTTPS clients are covered, ranging from niche web browsers to command line
tools and embedded libraries that are vital to protect but yet lack the
resources of major browser vendors~\cite{androidlibs,androidlibs2}.
Our approach coexists well with witness cosigning and cross-logging due to
different threat models, but not necessarily STH pushing if the secure
channel is encrypted (no need to fetch what a trusted party provides).

  \section{Discussion} \label{sec:discussion}
\fullversion{%
	Below we discuss assumptions, limitations, and deployment, showing that
	our approach towards retroactive gossip can be deployed at scale to detect
	split-views by many relevant attackers with relatively little effort.  The
	main drawback is reliance on clients fetching STHs in plaintext, e.g., using
	CT-over-DNS~\cite{ct-over-dns}.
}{}

\subsection{Assumptions and Limitations} \label{sec:discussion:limitations}
Aggregation-based gossip is limited to network traffic that packet processors
can observe. The strongest type of attacker in this setting---who can completely
isolate a client---trivially defeats our gossip mechanism and other retroactive
approaches in the literature (see Sect.~\ref{sec:related}).
A weaker attacker cannot isolate a client, but is located nearby in a network
path length sense. This limits the opportunity for packet processor aggregation,
but an attacker cannot rule it out given aggregation indistinguishability.
Sect.~\ref{sec:implementation} showed based on performance that it is non-%
trivial to distinguish between (non\mbox{-})aggregating packet processors on two
different targets using P4 and XDP. Off-path challengers must also be
indistinguishable from one another to achieve \emph{implicit gossip}.
While we suggested the use of anonymity networks like Tor, a prerequisite is
that this is in and of itself not an aggregation distinguisher.%
\footnote{%
	\fullversion{%
    Low-latency anonymity networks like Tor are susceptible to traffic
    confirmation and correlation attacks where the attacker observes traffic
    from the packet processor and is in control of the response from the CT
    logs. A strictly isolated packet processor may not be able to hide that it
	is challenging the logs (i.e., aggregation distinguisher).}
	{Low-latency anonymity networks are susceptible to traffic confirmation/%
	correlation. An isolated packet processor may not be able to hide that it is
	challenging the logs.}
}
Therefore, we assume that other entities also use off-paths to fetch and verify
STHs. The fact that a unique STH \emph{is not audited} from an off-path could
also be an aggregation distinguisher. To avoid this we could rely on a
verifiable STH history\footnote{%
	\fullversion
		{\url{https://web.archive.org/web/20170806160119/https://mailarchive.ietf.org/arch/msg/trans/JbFiwO90PjcYzXrEgh-Y7bFG5Fw} (2017)}
		{\url{https://mailarchive.ietf.org/arch/msg/trans/JbFiwO90PjcYzXrEgh-Y7bFG5Fw}}
}
and wait until the next MMD to audit or simply monitor the full log so that
consistency proofs are unnecessary.

The existence of multiple network paths are fundamental to the structure and
functioning of the Internet. A weak attacker may use IP fragmentation such that
each individual STH fragment is injected from a different location to make
aggregation harder, approaching the capabilities of a stronger attacker that
is located closer to the client. This is further exacerbated by the deployment
of multi-path transport protocols like MPTCP (which can also be fragmented).
Looking back at our RIPE Atlas measurements in Sect.~\ref{sec:measurements}, the
results towards Google's world-wide infrastructure better represent an active
attacker that takes \emph{some} measures to circumvent aggregation by
approaching a client nearby the edge. Given that the likelihood of aggregation
is high if \emph{any} IXP is present (Fig.~\ref{fig:wcov}), aggregation at
well-connected IXPs are most likely to be circumvented.

\subsection{Deployment} \label{sec:disussion:deployment}
Besides aggregating at strategic locations in the Internet's backbone,
ISPs and enterprise networks have the opportunity to protect all of their
clients with relatively little effort. Deployment of special-purpose middleboxes
are already prevalent in these environments, and then the inconvenience of
fragmentation tends to go away due to features such as packet reassembly.
Further, an attacker cannot trivially circumvent the edge of a network topology%
---especially not if aggregation takes place on an end-system:
  all fragments are needed to reassemble a packet, which means that multi-path
  fragmentation is no longer a threat.
If aggregation-based gossip is deployed on an end-system, STHs could be
hooked using other approaches than P4/XDP. For example, shim-layers that
intercept TLS certificates higher up in the networking stack were already
proposed by Bates~\emph{et~al.}~\cite{h1} and O'Neill~\emph{et~al.}~\cite{h2}.
In this setting an end-system is viewed as the aggregating packet processor,
and it reports back to an off-path challenger that may be a local process
running on the same system or a remote entity, e.g., a TelCo could host
challengers that collect aggregated STHs from smartphones.

While we looked at programming physical packet processors like routers,
STH aggregation could be approached in hypervisors and software switches~%
\cite{pisces} to protect many virtual hosts. If CT-over-DNS is used to fetch
STHs, it would be promising to output DNS server caches to implement the
aggregation step. Similar to DNS servers, so called Tor exist relays also
operate DNS caches. In other words, P4 and XDP are only examples of how to
\emph{instantiate} the aggregation step. Depending on the used plaintext source,
packet processor, and network topology other approaches may be more suitable,
e.g., C for vendor-specific middleboxes.

\subsection{Retroactive Gossip Benefits From Plaintext}
As opposed to an Internet core that only forwards IP packets, extra
functionality is often embedded which causes complex processing dependencies and
protocol ossification~\cite{TCPoss}. Many security and protocol issues were
found for middleboxes that provides extra functionality~\cite{HTTPSintercept,%
langely-quic}, resulting in the mindset that \emph{everything} should be
encrypted~\cite{langely-quic}.
Our work is controversial because it goes against this mindset and advocates
that STHs should be communicated in plaintext.
We argue that this makes sense in the context of STHs due to the absence of
privacy concerns and because the entire point of gossip is to make STHs
\emph{available} (rather than end-to-end only).
The idea of intentionally exposing information to the network is not new, e.g.,
MPQUIC is designed to support traffic shaping~\cite{mpquic}.

While we used CT-over-DNS as a plaintext source, there is a push towards
	DNS-over-TLS\footnote{%
		\fullversion
			{\url{https://web.archive.org/web/20180422194047/https://security.googleblog.com/2018/04/dns-over-tls-support-in-android-p.html} (2018)}
			{\url{https://security.googleblog.com/2018/04/dns-over-tls-support-in-android-p.html}}
	} and DNS-over-HTTPS\footnote{%
		\fullversion
			{\url{https://web.archive.org/web/20180512125541/https://blog.cloudflare.com/dns-resolver-1-1-1-1/} (2018)}
			{\url{https://blog.cloudflare.com/dns-resolver-1-1-1-1/}}
	}.
Wide use of these approaches could undermine our gossip mechanism, but
ironically the security of TLS could be jeopardized unless gossip is deployed.
In other words, long term gossip is an essential component of CT and other
transparency logs to avoid becoming yet another class of trusted third-parties.
If proactive approaches such as witness cosigning are rejected in favour of
retroactive mechanisms, then ensuring that STHs are widely spread and easily
accessible is vital. An STH needs no secrecy if the appropriate measures are
taken to make it privacy-insensitive~\cite{ietf-gossip}.
While secure channels also provide integrity and replay protection, an STH is
already signed by logs and freshness is covered by MMDs as well as issue
frequency to protect privacy.
A valid argument against exposing any plaintext to the network is protocol
ossification. We emphasize that our design motivates why packet processors
should fail open:
  otherwise there is no aggregation indistinguishability.

\subsection{Indistinguishability and Herd Immunity} \label{sec:discussion:herd}
An attacker that gains control over a CT log is bound to be more risk averse
than an attacker that compromises a CA. There is an order of magnitude fewer
logs than CAs, and client vendors are likely going to be exceptionally picky
when it comes to accepted and rejected logs.
We have already seen examples of this, including Google Chrome disqualifying
logs that made mistakes:
  Izenpe used the same key for production and testing,\footnote{%
	  \fullversion
		{\url{https://groups.google.com/a/chromium.org/forum/\#!topic/ct-policy/qOorKuhL1vA} (2016)}
		{\url{https://groups.google.com/a/chromium.org/forum/\#!topic/ct-policy/qOorKuhL1vA}}
} and Venafi suffered from an unfortunate power outage.\footnote{%
	\fullversion
		{\url{https://groups.google.com/a/chromium.org/forum/\#!topic/ct-policy/KMAcNT3asTQ} (2017)}
		{\url{https://groups.google.com/a/chromium.org/forum/\#!topic/ct-policy/KMAcNT3asTQ}}
}
Risk averse attackers combined with packet processors that are aggregation
indistinguishable may lead to \emph{herd immunity}: despite a significant
fraction of clients that lack aggregators, indirect protection may be provided
because the risk of eventual detection is unacceptable to many attackers. Hof
and Carle~\cite{hof-cross-logging} and Nordberg \emph{et~al.}~\cite{ietf-gossip}
discussed herd immunity~briefly~before~us.

  \section{Conclusion} \label{sec:conclusion}
Wide spread modifications of TLS clients are soon inevitable to support CT
gossip. We proposed that these modifications include challenging the logs to
prove certificate inclusion based on STHs \emph{fetched in plaintext}, thereby
enabling the traversed packet processors to assist in split view detection
retroactively by aggregating STHs for periodic off-path verification.
Beyond being an application neutral approach that is complementary to proactive
gossip, a compelling aspect is that core packet processors are used
  (rather than clients)
as a key building block to realize implicit gossip; should a consistency issue
arise, it is already in the hands of an entity that is well equipped to
investigate the cause manually.
Considering that far from all TLS clients are backed by big browser vendors---%
not to mention other use-cases of transparency logs in general---it is likely a
long-term win to avoid pushing complex retroactive gossip logic into all the
different types of clients when there are orders of magnitudes fewer packet
processors that could aggregate to their own off-path challengers.
While taking the risk of ossification into account by suggesting that packet
processors fail open to provide aggregation indistinguishability, our approach
offers rapid incremental deployment with high impact on a significant fraction
of Internet users.

  \ifx\doredact\undefined%
    \subsubsection{Acknowledgments.}
    We would like to thank Stefan Alfredsson and Philipp Winter for their RIPE Atlas
credits, as well as Jonas Karlsson and Ricardo Santos for helping with the
NetFPGA setup. We also received funding from the HITS research profile which is
funded by the Swedish Knowledge Foundation.

  \fi

  \bibliographystyle{splncs04}

  \appendix
  \section{Implementation} \label{app:implementation}
\fullversion{%
	Below implementation details are provided for CT-over-DNS as the aggregated
	plaintext source.  However, most of the discussion is not only relevant for
	DNS.
}{}

\subsection{Plaintext Source} \label{app:imp:pt}
Aggregation-based gossip relies on a plaintext source that packet processors
can observe.  The most applicable mechanism today is CT-over-DNS, which is
hosted by Google for all Chrome-included logs.  According to the draft by
Laurie~\cite{ct-over-dns}, a DNS STH response is an IN TXT resource record where
the query domain is
  \texttt{sth.<log>.ct.googleapis.com}.
We further restrict the format such that a response must be transported by UDP
and contain
   (i) a single query,
  (ii) a single response, and
 (iii) no more than a threshold of bytes.
It is hard to process a TCP data stream because it may span multiple IP packets.%
\footnote{%
	\fullversion
		{\url{https://web.archive.org/web/20180107232830/http://lists.p4.org/pipermail/p4-dev_lists.p4.org/2017-July/001176.html} (2017)}
		{\url{http://lists.p4.org/pipermail/p4-dev_lists.p4.org/2017-July/001176.html}}
}
At best, it is also hard to parse variable-length and human-readable protocols
such as HTTP.\footnote{%
	\fullversion
		{\url{http://web.archive.org/web/20190406091829/http://lists.p4.org/pipermail/p4-dev_lists.p4.org/2017-July/001175.html} (2017)}
		{\url{http://lists.p4.org/pipermail/p4-dev_lists.p4.org/2017-July/001175.html}}
}
From a general standpoint, this means that neither TLS $\le 1.2$ nor OCSP are
particularly prominent sources to aggregate despite STHs being transferred in
plaintext~\fullversion
	{\cite{ctga-thesis}}
	{\cite{ct}}.

\subsection{Proof-of-Concept} \label{app:imp:poc}
Figure~\ref{fig:parser} gives an overview of the headers that must be declared,
parsed, and inspected to aggregate DNS STHs. It is relatively easy to extract
headers down to DNS, after which the processing must continue in multiple
stages:
  extract a fixed-width preamble that contains the number of
    questions (\texttt{qd}) and
    answers (\texttt{an}),
  loop to extract the query domain name, and
  finally extract the fixed remainder of the query (type and class).
It is not possible to parse an arbitrary number of questions and answers in P4/%
XDP because loops must be constantly bound. This motivates the somewhat
restricted CT-over-DNS format in Appendix~\ref{app:imp:pt}.

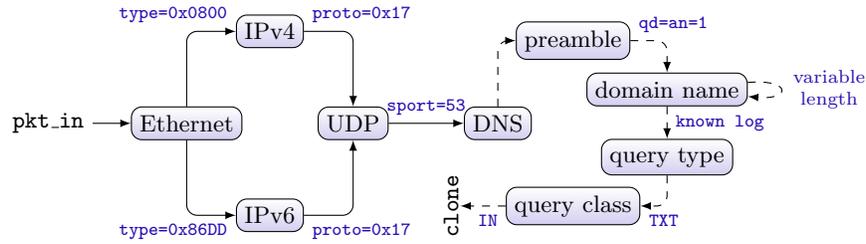
\begin{figure}[h]
  \ifx\dofullversion\defined\vspace{-.5cm}\fi
  \centering
  \begin{tikzpicture}[%
  -latex,
  sibling distance=10em,
  level distance=22pt,
  parser/.style = {%
    draw,
    shape=rectangle,
    rounded corners,
    align=center,
    top color=white,
    bottom color=mydblue!20,
  },
  label/.style = {%
    draw=none,
    align=center,
    text=mydblue,
    font=\scriptsize,
  },
  arrow/.style = {%
	  draw,
	  -latex,
	  rounded corners,
  },
]
	
	\node[parser](eth){Ethernet};
	\node[parser,right=of eth](udp){UDP};
	\coordinate(ip) at ($ (eth) !.5! (udp) $);
	\node[parser,above=of ip](ipv4){IPv4};
	\node[parser,below=of ip](ipv6){IPv6};

	\node[parser,right=of udp](dns){DNS};
	\node[parser](dnsp) at ($ (dns) + (1,1.1) $){preamble};
	\node[parser](dnsq) at ($ (dns) + (2.25,.45) $){domain name};
	\node[parser](dnst) at ($ (dns) + (2.25,-.45) $){query type};
	\node[parser](dnsc) at ($ (dns) + (1,-1.1) $){query class};

	\path[arrow] ($ (eth) + (-1.25,0) $) -- node[left,pos=0]{\texttt{pkt\_in}} (eth);

	\path[arrow] (eth) |- node[label,above left,pos=1]{\texttt{type=0x0800}} (ipv4);
	\path[arrow] (eth) |- node[label,below left,pos=1]{\texttt{type=0x86DD}} (ipv6);
	\path[arrow] (ipv4) -| node[label,above right, pos=0]{\texttt{proto=0x17}} (udp);
	\path[arrow] (ipv6) -| node[label,below right, pos=0]{\texttt{proto=0x17}} (udp);
	\path[arrow] (udp) -- node[label,above]{\texttt{sport=53}} (dns);

	\path[arrow,dashed] (dns) |- (dnsp);
	\path[arrow,dashed] (dnsp) -| node[label,above right,pos=0]{\texttt{qd=an=1}} (dnsq);
	\path[arrow,dashed] (dnsq) -- node[label,right]{\texttt{known log}} (dnst);
	\path[arrow,dashed] (dnst) |- node[label,below right,pos=1]{\texttt{TXT}} (dnsc);
	\path[arrow,dashed]
		(dnsq) edge[out=5, in=355, looseness=8]
			node[label,right]{
				\begin{tabular}{c}
					variable \\
					length
				  \end{tabular}
				}
		(dnsq);
	\path[arrow,dashed]
		(dnsc) --
			node[label,below right,pos=.8]{\texttt{IN}}
			node[pos=1.2,rotate=90]{\texttt{clone}}
		($ (dnsc) + (-1.5,0) $);
\end{tikzpicture}
  \caption{%
	  Criteria to aggregate an incoming packet \texttt{pkt\_in}.
  }
  \label{fig:parser}
  \ifx\dofullversion\defined\vspace{-.5cm}\fi
\end{figure}

If all conditions in Figure~\ref{fig:parser} hold for a packet, it is cloned
\emph{in addition to normal routing} using an existing P4-action or by
control-plane copying via XDP's ring-buffer\footnote{%
	\fullversion
		{\url{https://github.com/cilium/cilium/blob/master/Documentation/bpf.rst} (2018)}
		{\url{https://github.com/cilium/cilium/blob/master/Documentation/bpf.rst}}
}.  Note that small IP fragments which are less than a threshold are also marked
for cloning, accounting for attackers that intentionally fragment IP packets to
bypass aggregation.  Finally, our proof-of-concepts support simple probabilistic
filtering by cloning every $n^{\textrm{th}}$ match for a security parameter~$n$.

\subsection{Caveats} \label{app:imp:cav}
\subsubsection{Aggregation indistinguishability.}
Our packet processing is designed to avoid trivial aggregation distinguishers,
such as dropping tiny fragments proactively because they are more cumbersome to
validate (e.g., it requires reassembly).  Accordingly, it is paramount that
developers ensure that malformed packets are not dropped on parser exceptions
and that STH-related traffic remains unmodified by \emph{failing open}.  Given
that typical programs often operate on lower-layer headers, this is particularly
important while processing UDP and DNS headers.

\subsubsection{IP fragments and options.}
To minimize data collection an IP fragment should only be aggregated if it is
less than a threshold. Therefore, a log client must reject STH packets that are
too large.  At the time of writing a typical DNS STH is encoded as
$\approx$170~bytes, and a 400~byte threshold would presumably be large enough to
account for IP options, large domain names, and STH extensions (should they
exists in the future).
The privacy impact of aggregating small fragments appears to have little or no
impact on legitimate traffic~\cite{frag-study-02},\footnote{%
	\fullversion
		{\url{https://web.archive.org/web/20180612113649/https://tools.cisco.com/security/center/viewIpsSignature.x?signatureId=1206&} (2006)}
		{\url{https://tools.cisco.com/security/center/viewIpsSignature.x?signatureId=1206&}}
} which is reasonable given that the de-facto minimum MTU has been \emph{at
least} 576~bytes for decades~\cite{min-mtu,ipv6}. In other words, small
fragments are anomalies rather than expected behaviour. Under normal
circumstances and a sound STH frequency for privacy, we expect around 24 unique
STHs per day and log to be aggregated.

\section{Data Sets} \label{app:data}
\fullversion{%
	Our traceroute data set is publicly available and described in Appendix~%
	\ref{app:data:our}. The role of other public data sets used in our analysis
	are explained in Appendix~\ref{app:data:pub}.
}{}

\subsection{RIPE Atlas Traceroute Measurements} \label{app:data:our}
Our traceroute measurements can be downloaded from the RIPE Atlas
platform. Identifiers: \redact{%
  11603880--11603884,
  11784033--11784042, and
  11826645--11826649}.

\subsubsection{Probe selection.}
The goal of our probe selection process was to maximize the number of unique
ASes (which will represent blocks of IP addresses that we can evaluate coverage
for). The scope of our search was reduced to IPv4 because many probes support
it, and for redundancy the two most stable probes in each unique AS were
selected. We based the stable criteria on the RIPE Atlas tag
  \texttt{system-ipv4-stable-n},
such that a probe got the highest priority if $n{=}90$~days. While many ASes had
too few probes to support redundancy, we ended up requesting 4604~probes. After
removing the redundant probes that delivered the fewest amount of traceroute
results, there were little or no failures amongst the remaining 3512 (Google)
and 3488 (NORDUnet) probes:
  around 100 probes failed at least once, and among those
  24 as well as 17 probes (respectively) failed more than once.
This means that the reliability of RIPE Atlas platform is high, and thus it is
unnecessary to account for failures while analyzing our results.

\subsubsection{Duration and measurement settings.}
For all probes we scheduled a daily traceroute towards Google and NORDUnet.
Our measurements towards Google started on March~10 2018 and ended on March 30
2018. On March 20 we started another measurement towards NORDUnet that ended on
April~9 2018.  We used the RIPE Atlas default traceroute settings:
	ICMP port~80 with default spread and Paris traceroute enabled%
\fullversion
	{\footnote{\url{https://web.archive.org/web/20180511201452/https://paris-traceroute.net/} (n.d.)}}
	{}
(value 16).
The response timeout was set to 4000~ms for three 48~byte packets and 32~max
hops. We also hard-coded the targeted IP addresses because not all probes
support DNS lookups. To verify that the mapping from domain name to IP address
remained the same for Google's authoritative CT-over-DNS server, we conducted a daily santiy-%
check\footnote{%
  RIPE Atlas measurement identifiers: 11603871 and 11793938.
} from 128 worldwide probes that resolved
  \texttt{ctns.googleapis.com} $\equiv$ \texttt{216.239.34.64}
on the probes. An employee at SUNET verified that
  \texttt{plausible-fe1.ct.nordu.net} $\equiv$ \texttt{194.68.13.48}
would remain stable throughout the course of our experiments.

\subsection{Public Data Sets} \label{app:data:pub}
The traceroute data set in Appendix~\ref{app:data:our} contains lists of IP
addresses. Since we are interested in the actors that control the corresponding
packet processors, i.e., which actors are on a given path, we mapped each IP
address to an AS number and/or IXP
identifier using public data sets from Routeviews\footnote{%
  The Routeviews MRT format RIBs and UPDATEs Dataset,
  2018-03-12 14:00,
  \url{http://archive.routeviews.org/bgpdata/2018.03/RIBS/}
} and CAIDA\footnote{%
	\fullversion
		{The CAIDA UCSD IXPs Dataset, February 2018, \url{https://www.caida.org/data/ixps/}}
		{CAIDA UCSD IXPs Dataset, February 2018, \url{https://www.caida.org/data/ixps/}}
}.
We also relied on
  RIPE Atlas probe metadata to map probes to AS numbers,\footnote{%
    \url{https://atlas.ripe.net/docs/api/v2/reference/\#!/probes/probe\_list\_get}, accessed REST API 2018-04-06
  } CAIDA's largest AS rank to select globally influential ASes as aggregators,%
  \footnote{%
    \url{http://as-rank.caida.org/api/v1}, accessed REST API 2018-04-06
  } and Routeviews' data set to annotate each probe with the IPv4 space of its
  AS.

\end{document}